# Deep Transform: Cocktail Party Source Separation via Probabilistic Re-Synthesis

Andrew J.R. Simpson [#1]

[#] *Centre for Vision, Speech and Signal Processing, Surrey University*
*Surrey, UK*
[1] `Andrew.Simpson@Surrey.ac.uk`

*Abstract*—In cocktail party listening scenarios, the human brain is able to separate competing speech signals. However, the signal processing implemented by the brain to perform cocktail party listening is not well understood. Here, we trained two separate convolutive autoencoder deep neural networks (DNN) to separate monaural and binaural mixtures of two concurrent speech streams. We then used these DNNs as *convolutive deep transform* (CDT) devices to perform probabilistic re-synthesis. The CDTs operated directly in the time-domain. Our simulations demonstrate that very simple neural networks are capable of exploiting monaural and binaural information available in a cocktail party listening scenario.

*Index terms*—Deep learning, unsupervised learning, deep transform, probabilistic re-synthesis, source separation.

## I. Introduction

In cocktail party listening, a listener must selectively attend to a voice within a background of competing speech noise [1]. The fact that selective representations of speech emerge at a relatively early stage of the auditory pathway [2], [3] suggests that relatively simple bottom-up processes are involved. In addition, a role of synthesis is implied by the well known phenomenon of 'phoneme restoration' [4]. Therefore, a neural model capable of capturing both separation and synthesis in a bottom-up way may provide some insight into how the brain is able to perform cocktail party listening.

From a signal processing perspective, mixed sources may be separated using linear filters if they inhabit different regions of some abstract feature space. Hence, the primary problem is of finding an abstract transformation capable of projecting the competing signals into different regions of the same space. Once the signals have been transformed into an abstract feature space, in which they are separable, it is then necessary to filter the signals and invert the abstract transform so as to obtain the original signals. This may be interpreted as re-synthesis [5], [6].

One way to perform this abstract transformation is known as an autoencoder [7] or deep transform (DT) [5], [6]. The autoencoder DT is a deep neural network which is trained to replicate its inputs at its output layer. From the point of view of the autoencoder which is trained on the speech of a single speaker, source separation may be viewed as error correction wherein the speech of the competing speaker is treated as error [5]. However, viewing the autoencoder DT as an abstract filter, if the filter is trained to pass the speech of one speaker it will not be good at rejecting competing speech which exists within overlapping regions of the shared speech feature-space. Furthermore, in the training of the audoencoder on the speech of one speaker, there is no optimization constraint which forces the autoencoder to learn abstract representations that discriminate between different speakers.

To solve this problem, we introduced a discriminative, convolutive autoencoder DT which was trained to separate and re-synthesize speech from a mixture of two speakers. In order to maximize the ability of the autoencoder to learn discriminative filters, we trained it directly on both monaural and binaural time domain audio signals. The autoencoder learned to separate and re-synthesize the component monaural speech signals. We then used this convolutive DT to perform probabilistic separation and re-synthesis of the component speech.

## II. Method

We consider two equivalent simulated cocktail party listening scenarios, one monaural and one binaural, each involving two concurrent speakers (a man and a woman). Each speaker was separately recorded (in mono) reading from a different story. Both speech signals were adjusted to be of equal average intensity.

To produce a simulated monaural cocktail party listening scenario, the monaural speech from each speaker was linearly summed to produce a monaural mixture signal (see Fig. 1). To produce a simulated binaural cocktail party listening scenario, the monaural speech from each speaker was convolved with binaural head-related impulse responses (HRIR) corresponding to angles of incidence of +45 and -45 degrees (from center) respectively. This simulated two equidistant speakers at 90 degrees. The HRIR was obtained from a dummy head [8], recorded at 1m. A binaural mixture signal was obtained by summation of the two convolved speech signals (left and right channels were summed separately, giving left and right mixture channels, see Fig. 2).

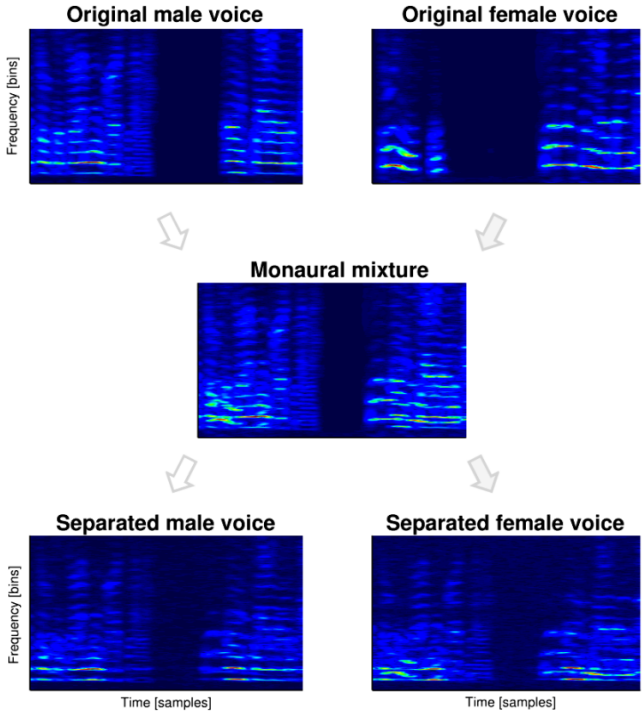

**Fig. 1. Monaural cocktail party source separation using convolutive DT probabilistic re-synthesis.** The upper pair of spectrograms plot a ~3-second excerpt from the original monaural audio for the male and female voice respectively. The middle spectrogram plots monaural mixture. The lower pair of spectrograms plot the respective CDT probabilistic re-synthesized and separated channels. This excerpt features (coincidentally) simultaneous utterances which result in overlapping vocalizations from both speakers in the mixture.

The mixture and original speech signals were decimated to a sample rate of 4 kHz. The first 2 minutes of the monaural speech signals and the corresponding monaural or binaural mixture were used as training data and the subsequent 10 seconds of audio was held back for later use in testing the separation of the models.

For training data, the monaural mixture, binaural mixture and the original monaural speech time domain signals were cut up into corresponding windows of 1000 samples (corresponding to a quarter of a second). The windows overlaped at intervals of 10 samples. Thus, for every quarter-second window, for training the monaural model there was a mixture signal (1000 samples) and two monaural input signals (1000 samples x 2), and for training the binaural model there was a two-channel binaural mixture signal (1000 samples x 2) and two monaural input signals (1000 samples x 2). This gave approximately 50,000 training examples (per model). For the testing stage, 10 seconds of speech/mixture was used at overlap intervals of 1 sample, giving approximately 40,000 test frames (which would ultimately be applied in an overlaping convolutional output stage). Prior to windowing, all audio was normalized to unit scale and mean of 0.5. This allowed the use of a signmoidal output function mapped to the range [0,1].

For the monaural model we used a feed-forward DNN of size 1000x2500x2000 units that was configured as an autoencoder such that the input layer was the mixture signal (1000 samples). For the binaural model we used a feed-forward DNN, of size 2000x2500x2000 units, that was configured as an autoencoder such that the input layer was the concatenated left and right channels (1000 + 1000 = 2000 samples). For both models, the output layer was trained to synthesize the concatenated male and female voice monaural inputs (1000 + 1000 = 2000 samples).

This discriminative arrangement of the output layer meant that the autoencoder was required to synthesize in the first 1000 output units the samples representing the monaural speech of the male voice and in the second 1000 output units it was required to synthesize the samples representing the monaural speech of the female voice. Thus, the binaural model was required to deconvolve (the HRIR) and unmix and re-synthesize both monaural signals that had been convolved with the binaural HRIR and mixed. Both the DNNs employed the biased-sigmoid activation function [9] throughout with zero bias for the output layer. Each autoencoder was trained using 300 full iterations of stochastic gradient descent (SGD). Each iteration of SGD featured a full sweep of the training data. Dropout was not used in training. After training, each autoencoder was used as a feed-forward signal processing device (i.e., a convolutive deep transform – CDT - [5], [6]).

*Probabilistic Re-Synthesis via CDT*. Each frame (length 1000 samples) of the test audio mixture was transformed using the respective autoencoder a number ($N$) of times. Prior to each of the $N$ separate transforms, 50% of the samples of the test frame (chosen at random) were replaced with random values of equivalent mean and standard deviation to the test data. I.e., the random perturbation of each frame, for each of the $N$ transforms, was independent. The activations of the first 1000 units (samples) of the output layer of the autoencoder were taken as re-synthesized audio for the male voice and the activations of the second 1000 units were taken as re-synthesized audio for the female voice. The resulting distribution, of $N$ re-synthesized instances of each perturbed audio frame, was then averaged to provide probabilistic estimates of the true input frames. In order to account for neurons in the output layer that were invariantly active, all estimated (separated) frames (i.e., of the entire test set) were then averaged and the result subtracted from each frame. The frames were then superposed in a sliding-window fashion, the results averaged and the DC offset removed (zero mean was restored).

Separation quality (for the test data) was measured using the BSS-EVAL toolbox [10] and is quantified in terms of signal-to-distortion ratio (SDR), signal-to-artefact ratio (SAR) and signal-to-interference ratio (SIR).

## III. RESULTS

Fig. 1 plots spectrograms illustrating the stages of monaural mixture and separation for a brief (~3 second) excerpt from the test data. Fig. 2 illustrates the equivalent for the binaural condition (for the same speech excerpt from the test data). In both figures, by way of benchmark for separation, the spectrograms for the monaural speech signals are shown at the top. The middle panel of Fig. 1 plots the monaural mixture and the middle panel of Fig. 2 plots the HRIR-convolved binaural (left and right) spectrograms. The mixture spectrograms illustrate coincidental overlap of the two competing speech signals in time-frequency space. At the bottom of Fig. 1 are plotted the separated and re-synthesized audio for the male and female voices respectively (resampling rate of $N = 100$). At the bottom of Fig. 2 are plotted the equivalent separated and re-synthesized audio from the binaural model ($N = 100$).

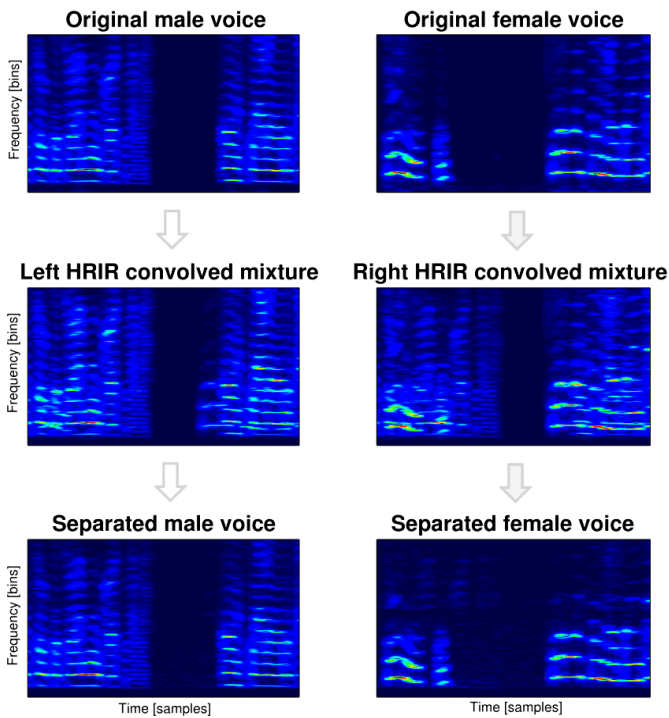

In the excerpt illustrated here there are strongly overlapping vocal (tonal) components (i.e., partials). The monaural model (Fig. 1) appears to provide some (but not ideal) selectivity for the male/female voices respectively. In particular, the lower frequency overlaping partials appear to have been separated relatively well. By contrast, the binaural model (Fig. 2) shows practically ideal separation of the tonal signal components. According to an informal audition, the separated and re-synthesized audio is of good quality and some selectivity is audible in the monaural case. In the binaural case, the selectivity and overall quality is good.

Fig. 3 plots the various objective source separation quality metrics (SDR/SIR/SAR), computed over the entire 10-second test data, as a function of resampling rate (*N*). In general, the measures tend towards asymptotic at around $N=100$ illustrating the convergence of the probabilistic re-synthesis. In the monaural case (Fig. 3a), there is a strong disparity between the results for the male voice (solid lines) and the female voice (dashed lines). This is broadly consistent with what is shown in Fig. 1, where it appears that suppression of the female voice in the male-selective output channel is more successful than vice-versa. The binaural results (Fig. 3b) are better than in the monaural case. This most likely reflects the fact that the binaural mixture (for either left or right channel) featured less equal energy ratios. It also presumably reflects the fact that binaural cues are far more useful in separating the speech than monaural spectro-temporal cues. In the binaural case (Fig. 3b), there is a small disparity (favouring the male voice) that appears broadly equivalent to that of the monaural model, but overall the separation results are much more balanced. Thus, it seems likely that the unequal selectivity observed in the monaural model may be due to some possible energetic asymmetry which acts (through the cost function) during training to bias the model towards the male voice.

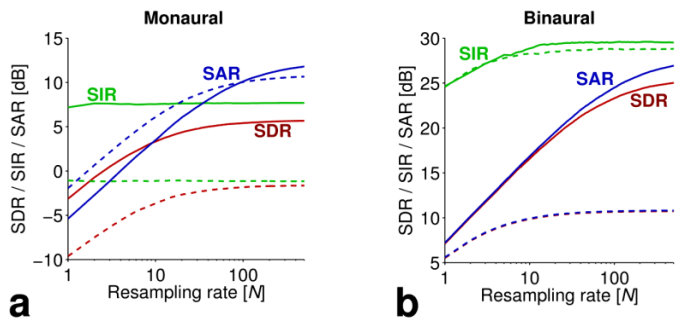

**Fig. 2. Binaural cocktail party source separation using convolutive DT probabilistic re-synthesis.** The upper pair of spectrograms plot a ~3-second excerpt from the original monaural audio for the male and female voice test data respectively. The middle spectrograms plot the respective left and right binaural mixtures (convolved with the respective HRIRs). The lower pair of spectrograms plot the respective CDT probabilistic re-synthesized and separated channels. As in Fig. 1, this excerpt features (coincidentally) simultaneous utterances which result in overlapping vocalizations from both speakers in the mixture.

**Fig. 3. Monaural and binaural separation quality measures.** Signal-to-distortion ratio (SDR, red), signal-to-interference (SIR, green), signal-to-artefact ration (SAR, blue), computed from the 10-second test audio using the BSS-EVAL toolkit [10]. **a** plots SDR/SIR/SAR as a function of resampling rate (*N*) for the male voice (solid lines) and female voice (dashed lines) respectively. **b** plots the same for the binaural equivalent.

## IV. DISCUSSION AND CONCLUSION

We have demonstrated that CDT probabilistic re-synthesis [6] may be applied to the problem of source separation directly in the time domain. We have also contrasted the difficulty of the problem for monaural and binaural cocktail party listening scenarios. It appears that the monaural model has learned those spectro-temporal features which discriminate the voices, whereas the binaural model has learned the HRIR localization features (cues) which more easily discriminate the angles of incidence of the voices.

While the monaural model provides relatively poor selectivity at least for the female voice (SIR – Fig. 3a), the measures reflecting sound quality (SDR/SAR) appear to reflect the fact that these models operate in the time domain, where there are no issues with missing or corrupted phase information and no need for inverse Fourier transform. This work may be applicable to hearing aid technologies and other separation problems where time-domain solutions may be useful.

More generally, in the human auditory system binaural integration occurs early (at the level of the brainstem). The marked advantage of the binaural separation model demonstrated here tends to suggest that the brain exploits similar strategies for separation and may begin the separation process at such early stations. Hence, our findings may provide some insight into the architecture and function of the auditory brain in cocktail party listening scenarios.

ACKNOWLEDGMENT

AJRS was supported by grant EP/L027119/1 from the UK Engineering and Physical Sciences Research Council (EPSRC).